\documentclass[a4paper,11pt,reqno]{amsart}
\newtheorem{theorem}{Theorem}
\newtheorem{proposition}{Proposition}
\newtheorem{lemma}{Lemma}
\newtheorem{corollary}{Corollary}

\newtheorem{remark}{Remark}

\def\L{{\mathcal L}}
\def\B{{\mathcal B}}

\def\H{{\mathcal H}}

\def\R{{\mathcal R}}

\def\G{{\mathcal G}}
\def\S{{\mathcal S}}

\def\miolemma#1#2{\begin{lemma}\label{#1}#2\end{lemma}}

\def\proof{\noindent{\bf Proof.\ }}
\def\quadratino{
\hfill\vbox{\hrule\hbox{\vrule\vbox to 7 pt {\vfill\hbox to
7 pt {\hfill\hfill}\vfill}\vrule}\hrule}\par}

\def\Za{{\mathbb Z}}
\def\Re{{\mathbb R}}

\def\Cm{{\mathbb C}}
\def\toro{{\mathbb T}}

\def\perogni{\forall\hskip1pt}

\def\norma#1{\left\Vert#1\right\Vert}

\def\normaop#1#2{\norma{#1}_{#2}}
\def\normalip#1{\norma{#1}^{\L}}

\def\i{{\rm i}}
\def\e{{\rm e}}
\def\sleq{\leq\kern-4pt\cdot}
\def\sgeq{\null\kern+1pt\cdot\kern-4.5pt\geq}
\def\om{\omega}
\def\be{\begin{equation}}
\def\ee{\end{equation}}

\begin{document}
\baselineskip=18pt

\title[Nonautonomous Schr\"odinger operators and KAM
methods]{Time Quasi-periodic unbounded perturbations of Schr\"odinger
operators and KAM methods }
\author{ Dario Bambusi $\quad$ Sandro Graffi}
 \address{ Dipartimento di
Matematica ``F. Enriques'', Universit\`a di Milano \newline \hphantom{vi} Via
Saldini 50, 20133 Milano, Italy.}
\email{bambusi@mat.unimi.it}
\address{Dipartimento di Matematica,  Universit\`a di Bologna  \newline
\hphantom{vi} Piazza di Porta S Donato 5, 40127 Bologna, Italy.} 
\email{graffi@dm.unibo.it} 

\begin{abstract}
We eliminate by KAM methods the time dependence in a class of linear
differential equations in
$\ell^2$ subject to an unbounded, quasi-periodic forcing.  This entails 
the pure-point nature of the Floquet spectrum of the operator  $
H_0+\epsilon P(\om t)$ for $\epsilon$ small.  Here $H_0$ is the
one-dimensional Schr\"odinger operator
$p^2+V$, $V(x)\sim |x|^{\alpha}, \alpha >2$ for
$|x|\to\infty$, the time quasi--periodic perturbation $P$ may grow as
$\displaystyle |x|^{\beta}, \beta <(\alpha-2)/{2}$, and the frequency
vector $\omega$ is  non resonant. The proof 
extends to infinite dimensional spaces the  result valid for
quasiperiodically forced linear differential equations and is based on
Kuksin's estimate of solutions of homological equations with non
constant coefficients.
\end{abstract}

\maketitle


\section{Introduction and statement of the results}
\setcounter{equation}{0}%
\setcounter{theorem}{0}%
\setcounter{proposition}{0}%
\setcounter{lemma}{0}%
\setcounter{corollary}{0}%
\setcounter{definition}{0}%
\setcounter{remark}{0}%
\setcounter{example}{0}%
\noindent
Consider the   non-autonomous,  linear 
differential equation in a se\-parable Hilbert space ${\mathcal H}$
\be
\label{I0}
\i\dot{\psi}(t)=(A+\epsilon P(\omega_1 t,\omega_2 t,...,\omega_n
t))\psi(t) ,\quad \psi(t)\in {\mathcal H}, \;\epsilon\in\Re
\end{equation}
under the following conditions:
\begin{itemize}
\item[A1]  The operator $A$ is  positive self-adjoint. ${\rm Spec}(A)$ is
discrete, and all eigenvalues
  $0<\lambda_1<\lambda_2<\lambda_3,...$ are simple. There
is
$d>1$ such that 
\begin{equation}
\label{asy1}
\lambda_i\sim i^d\ ,\quad i\to\infty .
\end{equation}
\item[A2] $P(\phi_1,...,\phi_n)\equiv P(\phi)$ is a function from the
$n$-dimensional torus $\toro^n\equiv \Re^n/2\pi Z^n$ into the
symmetric operators in ${\mathcal H}$, $\omega:=
(\om_1,\ldots,\om_n)\in [0,1]^n$ is a frequency vector. 
\item[A3] For $\delta\geq 0$, denote $\B^{\delta}$  the Banach space of
all closed operators  $T$ in ${\mathcal H}$ such that
$A^{-\delta/d}T$ is bounded (remark that $\B^0={\mathcal L}({\mathcal
H})$), with norm
\begin{equation}
\label{norde}
\|T\|_{\delta}:=\sup_{\norma x_{{\mathcal H}}=1}\|
A^{-\delta/d}Tx\|_{{\mathcal H}}
\end{equation}

Then the map $\displaystyle \toro^n\ni\phi\to
P(\phi)\in\B^\delta$
is analytic for some $\delta<d-1$.
\end{itemize} 
Our purpose is to prove the following
\begin{theorem}
\label{main}
There exist $\epsilon_* >0$, 
 a
subset $\Pi^\epsilon\subset\Pi:=[0,1]^n$ and, if
$|\epsilon|<\epsilon_*$ and  $\omega\in\Pi^\epsilon$, 
a unitary operator $U_\epsilon(\omega t)\equiv U_\epsilon(\omega_1 t,\omega_2t,...,\omega_nt) $ in ${\mathcal H}$ with the
following  properties:
\begin{itemize}
\item[{T1}] $U_\epsilon(\omega t)$   
 is analytic in $t$ and quasiperiodic
 with frequencies  $\omega$;
\item[{T2}]  $U_\epsilon(\omega t)$  
  transforms equation (\ref{I0})
into a  system of the form
\begin{eqnarray}
\label{diagonale}
& \i \dot{\chi}(t) =A_{\infty}(\omega t){\chi}(t)
\\
& A_{\infty}:=
{\rm diag}(\lambda^\infty_1+\mu_1^\infty (\omega
t),\lambda^\infty_2+\mu_2^\infty(\omega t),\lambda^\infty_3
+\mu_3^\infty(\omega t),...)\ 
\end{eqnarray}
Here $\{\lambda^\infty_i\}_{i=1}^{\infty} \in\Re$ and any function $
\mu_i^{\infty}(\phi):\toro^n\to\Re$ is analytic with zero average;
\item[T3] There exists $C>0$ such that:
$$
\|1-U_{\epsilon}(\omega t)\|_0\leq C \epsilon \ ,\quad
|\lambda^\infty_i-\lambda_i|\leq C i^{\delta}\epsilon\ ,\quad
|\mu_i(\omega t)|\leq C i^{\delta}\epsilon\ ,\quad
\left|\Pi-\Pi^\epsilon\right| \mathop{\to}^{\epsilon\to0}0\ .
$$
\end{itemize}
\end{theorem}
Straightforward integration of (\ref{diagonale})  reduces (\ref{I0}) 
 to an autonomous system which makes the almost-periodic nature of
all its solutions evident.
\begin{corollary}
\label{cor1}
\begin{enumerate}
\item If $|\epsilon|<\epsilon_*$, $\omega\in\Pi^\epsilon$
there
exists a unitary transformation 
$U_F(\omega t)$, qua\-si\-periodic 
with frequency $\omega$  and such that
$\norma{1-U_F(\omega t)}_{\delta}\leq C \epsilon$,  which
transforms  (\ref{I0}) into the system
\begin{equation}
\label{diag}
\i \dot x =A_{F}x, \quad 
A_{F}:={\rm
diag}(\lambda^{\infty}_1,\lambda^{\infty}_2,\lambda^{\infty}_3,...)\ ;
\end{equation}
\item
For any initial datum  $\;\psi_0$ the solution $\;\psi(t)$ of (\ref{I0})
is almost-periodic with frequencies $2\pi/\lambda^\infty_1,2\pi/
\lambda^\infty_2, \ldots; \om_1,\ldots,\om_n$, i.e. has the form
\begin{equation}
\label{psi}
\psi(t)=\sum_{i=0}^{\infty}\phi_i^0(\om t)e^{i\lambda^{\infty}_i t}
\end{equation}
where $\{\phi_i^0(\om t)\}_{i=1}^{\infty}$ are the 
components of $U_{\epsilon}(\om t)\psi_0$ along the eigenvector basis of $A$.
\end{enumerate}
\end{corollary}
The above result can be equivalently formulated in terms of 
Floquet spectrum (\cite{Ya}, and \cite{Le} for the quasi-periodic case).
Consider indeed on
${\mathcal K}:=\H\otimes L^2(\toro^n)$
 the  Floquet Hamiltonian operator
\begin{equation}
\label{flo}
K_F:=-\i \sum_{l=1}^n\omega_l\frac\partial{\partial \phi_l} +A+\epsilon
P(\phi)\ .
\end{equation}
The maximal operator in ${\mathcal K}$ generated by the
differential expression (\ref{flo}), still denoted 
$K_F$, is  self-adjoint  by A3, which makes $A+\epsilon P(\om t)$
self-adjoint on $D(A)$ for all $t$. Then:
\begin{corollary} 
\label{mainc}
For 
$|\epsilon|\leq \epsilon_*$ and $\om\in\Pi^\epsilon$ the spectrum of
$K_F$  is pure point;
its eigenvalues are $\nu_{j,k}:=\lambda_j^{\infty} +
k\cdot\omega$, $j=0,1,2\ldots, $ 
$k\in\Za^n$.
\end{corollary}
\begin{remark}
\label{Oss1}
\begin{enumerate}
{\rm \item This corollary extends to  unbounded and quasiperiodic
perturbations  the analogous result valid for operators
$K_F$ with $P(\phi)$ periodic and differentiable
in $\phi$ as a bounded operator in ${\mathcal H}$ \cite{DuS,DuSV}. The
KAM methods of \cite{DuS,DuSV}, first implemented  in 
\cite{Be} (see also \cite{Co}) made possible to strengthen for small
coupling the original result of
\cite{Ho} (see also
\cite{Jo},\cite{Ne}) from absence of absolutely continuous spectrum to
absence of continuous spectrum. Here too the set $\Pi^\epsilon$ is the
set of all frequencies fulfilling a diophantine condition with respect
to the differences $\lambda_i -\lambda_j$. Moreover, a
 result of the type of Corollary \ref{cor1} up to an
error of order $\displaystyle \exp{1/\epsilon*}$ has been proved in
\cite{JM} for a class of bounded perturbations via the Nekhoroshev
technique.
 \item
Our proof extends to infinite dimensional spaces the KAM 
technique  to e\-li\-mi\-na\-te the time dependence of
quasiperiodically forced ordinary linear differential equations
\cite{arnold,spagnoli,cinesi}.  The main technical point is that the
relevant homological equation has variable coefficients but can be solved
by a technique developed by Kuksin\cite{kuk97} in the context of his
analysis of the KdV equation by KAM theory.
} 
\end{enumerate}
\end{remark}
As in \cite{Co, DuS,DuSV,Ho,Jo,Ne,JM}  the main  motivation
for this corollary  is the  (Floquet) spectral analysis for the time
dependent Schr\"o\-dinger equation in dimension one, namely:
\begin{theorem}
\label{Floquet}
Consider  the time dependent Schr\"o\-dinger equation 
\begin{equation}
\label{Sc}
 H(t)\psi(x,t)=i\partial_t \psi(x,t),\;x\in\Re;\quad
 H(t):=-\frac{d^2}{dx^2} +Q(x)+\epsilon V(x,\omega
t),\; \epsilon\in\Re 
\end{equation}
and the corresponding Floquet Hamiltonian (\ref{flo})
under the following conditions:
\begin{enumerate}
\item $Q(x)\in C^{\infty}(\Re;\Re)$, $Q(x)\sim |x|^{\alpha}$ for some $\alpha
>2$ as $|x|\to \infty$;
\item
$V(x,\phi)$ is a $C^{\infty}(\Re;\Re)$-valued holomorphic function of
$\phi\in\toro^n$, with $\displaystyle |V(x,\phi)| |x|^{-\beta}$ boun\-ded as 
$|x|\to
\infty$ for some $ \displaystyle\beta <  \frac{\alpha -2}{2}$.

\end{enumerate}
Then there is $\epsilon^* >0$ such that the spectrum of $K_F$ is pure
point for all $|\epsilon|<\epsilon^* $, $\om\in \Pi^\epsilon$.
\end{theorem}
\noindent
\begin{remark}
\begin{enumerate}
{\rm \item
 We  prove the result in the more general case where  
$V$   is a  $C^{\infty}(\Re^2;\Re)$-valued holomorphic function
$V(x,\xi;\phi)$ of
$\phi\in\toro^n$ with $\displaystyle
|V(x,\xi;\phi)|(|\xi|^2+|x|^{\alpha})^{-\delta/d}$ boun\-ded as
$|\xi|+|x|\to\infty$. Here  $V(\phi)$ is realized as a
pseudodifferential operator family in $L^2(\Re)$ of class $G^{\beta}_{\rho}$
(see e.g.\cite{Sh}, Chapter 8) of Weyl symbol $V$.
\item 
For $\alpha=4$ we get $\beta <1$. Hence the quantum version of
the original  Duffing oscillator $\displaystyle
H(t)=-\frac{d^2}{dx^2}+x^4+\epsilon
x\sin{(\om t)}$  lies just outside the validity range of this corollary.
\item In the periodic case $(n=1)$ 
 we see  that, as in classical mechanics (see e.g.\cite{Ga}, Chapt.5.13) not
even an unbounded perturbation delocalizes the system if its strength
$\epsilon$ is too small and its frequency is not too close to a resonant one. 
 There is no diffusion (for $\epsilon$
small enough)  in the classical counterpart of (\ref{Sc}) even for resonant
values of $\omega$, but there are chaotic regions in phase space
localized around the resonant actions. In this case  it is still unkown
whether or not  the quantum Floquet spectrum is  pure point  even for bounded
perturbations. 
On the other hand for $0<\alpha \leq 2$, when condition (\ref{asy1}) is
not satisfied, the nature of the Floquet spectrum is still unknown  apart
the globally resonant case\cite{GY},\cite{HSl}.
\item In the quasiperiodic case ($n\geq 2$) the quantized system behaves 
 as
in the periodic one even though in the classical
counterpart of (\ref{Sc}) there are no topological obstructions to the
growth of energy.}
\end{enumerate}
\end{remark}

\section{The formal construction}
\setcounter{equation}{0}%
\setcounter{theorem}{0}%
\setcounter{proposition}{0}%
\setcounter{lemma}{0}%
\setcounter{corollary}{0}%
\setcounter{definition}{0}%
\setcounter{remark}{0}%
\setcounter{example}{0}%

Without loss of generality equation (\ref{I0}) can be written as a
first-order system in $\ell^2$:
\begin{eqnarray}
\label{eq0}
\i   \dot x=(A+\epsilon P(\omega t))x \ ,\qquad  x\in\ell^2 \qquad
\\
\label{A0}
A={\rm diag}(\lambda_1,\lambda_2,\lambda_3,...)\ ,\quad \lambda_i\in\Re\
,\quad \lambda_i>0
\end{eqnarray}
where $\lambda_i$ and $P(\omega t)\equiv P(\omega_1
t,\omega_2t,...,\omega_n t)$ fulfill conditions A1-A3.

The key point of any KAM method is the construction of a coordinate
transformation mapping the original problem into a new one of the same
form with a much smaller size of the perturbation, typically the
square of the original one. Here we construct and estimate, by an
algorithm very close to that of \cite{JM}, a unitary operator which
maps (\ref{eq0}) into an equation of the same form but with a
perturbation of order $\epsilon^2$. 

In this Section we describe the procedure; in Sect. 3 we work out the
estimates, and in Sect.4 we set up the iterative scheme and prove its
convergence.

Let 
$B(\phi_1,...,\phi_n)\in\B^0$
be an\-ti-self\-ad\-joint
 $\forall\,\phi\in\toro^n$. Given the unitary
operator
$e^{\epsilon B(\phi)}$,  for fixed  $\omega\in \Pi$ perform the
  change of basis $x=e^{\epsilon B(\omega t)}y$.  Substitution in 
 (\ref{eq0}) yields
\begin{eqnarray}
\label{eqy}
  \i \dot y=(A+\tilde P^1(\omega t))y
\end{eqnarray}
The new perturbation $\tilde P^1$ is  (the
explicit dependence of $B$ on $t$ is omitted):
\begin{equation}
\label{P1}
\begin{split}
& \tilde P^1:=\epsilon \left\{[A,B]-\i\dot B +P\right\}  
\\
&
+\left(e^{-\epsilon B}Ae^{\epsilon B}-A-\epsilon[A,B]
\right)
+\epsilon \left(e^{-\epsilon B}P e^{\epsilon B}-P \right)
-\i\epsilon \left(e^{-\epsilon B}\dot Be^{\epsilon B}-\dot B\right).
\end{split}
\end{equation}
If $B$  makes the curly bracket
vanish $\tilde P^1$ becomes of order $\epsilon^2$. Hence we  study
the equation 
\begin{equation}
\label{hom0}
[A,B]-\i\dot B +P=0\ .
\end{equation}
Taking its matrix  elements between the eigenvectors of $A$ this equation
becomes
\begin{equation}
\label{comp0}
-\i\sum_{l=1}^{n}\omega_l\frac{\partial}{\partial \phi_l} B_{ij}+
(\lambda_i-\lambda_j)
B_{ij}=P_{ij}\ ,
\end{equation}
Expand both sides in  Fourier series, i.e. write 
$$
 B_{ij}=\sum_{k\in Z^n}\hat B_{ijk}\e^{\i k\cdot\phi}, \qquad
P_{ij}=\sum_{k\in Z^n}\hat P_{ijk}\e^{\i k\cdot\phi}.
$$
 Equating the
Fourier coefficients of both sides  (\ref{comp0}) becomes
$$
(\omega\cdot k+\lambda_i-\lambda_j)\hat B_{ijk}=\hat P_{ijk}\ .
$$
Clearly this equation cannot be solved when $i=j$ and $k=0$. Assuming now
 $\omega$  such that
$\omega\cdot k+\lambda_i-\lambda_j\not=0$ when $i\not=j$ or $k\not=0$,
the natural definition of $B$ would be the operator with matrix elements
defined as
\begin{equation}
\label{pro}
\begin{split}
B_{ij}:=\sum_{k\in Z^n}\frac{\hat P_{ijk}}{\omega\cdot
k+\lambda_i-\lambda_j} \e^{\i k\cdot\phi}\ , \quad i\not =j \\
B_{ii}:=\sum_{k\in Z^n-\{0\}}\frac{\hat P_{iik}}{\omega\cdot k}\e^{\i
k\cdot\phi}\
\end{split}
\end{equation} 
The second line in
(\ref{P1}) is of order $\epsilon^2$ only if the operator $B$ is
bounded. However $P$ is not bounded; as a consequence the
operator
${\rm diag}(B_{ii})$ is in general unbounded, and the above
definition cannot yield the desired result. The idea is therefore to
define $B$ by the first of (\ref{pro}) with 
$B_{ii}=0$; 
one can guess that, since the
denominators $\omega\cdot k+\lambda_i-\lambda_j $ tend to
infinity as $i$ or $j$ diverge, it should be possible to generate a
bounded $B$ even if $P$ is unbounded. In the next section we will
prove that this is actually the case. 

With the above definition of $B$
the curly bracket in (\ref{P1}) turns out to be the
operator $\epsilon\, {\rm diag}(P_{ii})$, and hence in terms
of the variables $y$ the equation  takes the form. 
$$
\i\dot y= (A^1+\epsilon^2 P^1(\omega t))y\ ,
$$
with $A^1=A+\epsilon {\rm diag}(P_{ii}(\omega t))$. This
system is defined only for $\omega$ in the subset of $\Pi$ where the
denominators in (\ref{pro}) do not vanish. In the next section we will
assume  a diophantine type condition also for such denominators, to be
valid on a Cantor subset of $\Pi$. Then it will turn out that
$P^1$ depends in a Lipschitz way on $\omega$ in such a subset.

Iterating the construction, we see that the operator $A$ is replaced by
the operator $A^1$ which depends also on the angles $\phi$. As we shall
see, this is precisely the point where Kuksin's result\cite{kuk97} enters
in a critical way.

\section{Squaring the order of the perturbation}
\setcounter{equation}{0}%
\setcounter{theorem}{0}%
\setcounter{proposition}{0}%
\setcounter{lemma}{0}%
\setcounter{corollary}{0}%
\setcounter{definition}{0}%
\setcounter{remark}{0}%
\setcounter{example}{0}%

Keeping in mind the discussion of the preceding section we first set 
some notation, and then construct and estimate the transformation squaring
the order of the perturbation.

Let $\toro^n_s$  be the
complexified torus with $\left|{\rm Im}\phi_i\right|\leq
s$.  If $f$ is an analytic function from $\toro^n_s$ to a Banach
space (in what follows $\Cm$ or the complexification of $\B^\delta$),
we  denote
$$
\norma{f}_{s}=\sup_{\phi\in\toro^n_s}\norma{f(\phi)}
$$
For  $\B^{\delta}$-valued functions we  use the
particular symbol
$$
\|f\|_{\delta,s}:=\sup_{\phi\in\toro^n_s}\normaop{f(\phi)
}\delta\ 
.
$$ 
Let $\Pi^-$ be a closed nonempty subset of $\Pi$ of positive measure. If
$f$ has an additional  (Lipschitz continuous) dependence on
$\omega\in\Pi^-$ we define the norm
$$
\normalip{f}_s:=\norma{f}_s+\sup_{\phi\in\toro^n}
\sup_{\omega,\omega'\in\Pi^-}\frac{\norma{f(\phi,\omega)- 
f(\phi,\omega')}}{|{\omega-\omega'}|} \ .
$$
In particular for $\B^\delta$-valued functions  we 
use the notation $\normalip._{\delta,s}$. 

\vskip 5pt

Let us now include our system into a more general framework, which, by
the above discussion, is convenient for the iteration scheme.
Consider in $\ell^2$ the equation
\begin{equation}
\label{eq}
\i  \dot x=(A^-+P^-(\omega t))x
\end{equation}
under the following conditions

\begin{itemize}
\item[H1)]
\begin{equation}
\label{A}
A^-={\rm diag}(\lambda^-_1(\om)+\mu_1^-(\omega
t,\om),\lambda^-_2(\om)+\mu_2^-(\omega t,\om),\lambda^-_3(\om)+\mu_3^-(\omega
t,\om),...)\ ,
\end{equation}
Here:

\item[H1.a)] $\forall i=1,\ldots$ $\lambda_i^-(\om)$ is 
 positive and Lipschitz continuous w.r.t. $\omega\in\Pi^-$; moreover 
$$
 \lambda_i^-\sim i^d\ ,\quad 
$$
uniformly in  $\omega\in\Pi^-$. Hence there is 
 $C^-_\lambda >0$ independent of $\om$ such that
\begin{equation}
\label{clambda}
\left|\lambda^-_i-\lambda^-_j\right|\geq C^-_\lambda|i^d-j^d|\ .
\end{equation}
\item[H1.b)] There is  $C_\omega^- >0$ suitably small and $\delta<d-1$ such
that
\begin{equation}
\label{lip}
\sup_{\omega,\omega'\in\Pi^-}
\frac{|\lambda^-_i(\omega)-\lambda^-_i(\omega')|}{|\omega-\omega'|}\leq
C_\omega^-i^\delta 
\end{equation}
\item[H1.c)] $\forall i=1,\ldots$ $\mu_i^-(\om):\toro^n_s\times\Pi^-\to\R$ is
analytic w.r.t. $\phi$,  Lipschitz continuous  w.r.t.
$\omega$, and has zero average, i.e. 
$$
\int_{\toro^n}\mu_i(\phi,\omega)\,d\phi=0 .
$$ 
Moreover it fulfills the  estimates
\begin{equation}
\label{dio3}
\norma{\mu_i}_s\leq C^-_{\mu}i^\delta
\end{equation}
\begin{equation}
\label{lipmu}
\sup_{\phi\in\toro^n_s}\sup_{\omega,\omega'\in\Pi^-}
\frac{|\mu^-_i(\omega,\phi)-\mu^-_i(\omega',\phi)|}{|\omega-\omega'|}\leq
C_\omega^-i^\delta 
\end{equation}

\item[H2)] The operator valued function
$P^-:\toro^n_s\times\Pi^-\to\B^{\delta}$ is analytic with respect to
$\phi\in\toro^n_s$ and Lipschitz continuous w.r.t. $\omega\in\Pi^-$.

\item[H3)] there exist $\gamma^->0$ and $\tau>n+2/(d-1)$ such that,
for any $\omega\in\Pi^-$, one has 
\begin{equation}
\label{dio1}
|\omega\cdot k|\geq\frac{\gamma^-}{|k|^{\tau}}\ ,\quad \perogni
k\in\Za^n-\{0\}\ ,
\end{equation}
\begin{equation}
\label{dio2}
|\lambda_i-\lambda_j+\omega\cdot
 k|\geq\frac{\gamma^-|i^d-j^d|}{1+|k|^{\tau}}\ , \perogni k\in\Za^n\
 ,\quad i\not=j
\end{equation}
\end{itemize}
\vskip 0.3cm

\begin{remark}{\rm
In the next section we will prove that it is possible to construct a
set $\Pi^-$ of positive measure such that also the original system
(\ref{I0}) fulfills the above assumption.}
\end{remark}

Let now
\begin{equation}
\label{B}
  B:\toro_s^n\ni(\phi_1,...,\phi_n)\mapsto B(\phi_1,...,\phi_n)\in\B^0
\end{equation}
be an analytic map with $B(\phi_1,...,\phi_n)$ 
anti-selfad\-joint for each real value of
$(\phi_1,...,\phi_n)$. Consider the corresponding unitary operator
$e^{B(\phi_1,...,\phi_n)}$, and (as above) for any $\omega\in \Pi^-$
consider the unitary change of basis $x=e^{B(\omega t)}y$.
Substitution in equation \ref{eq} yields
\begin{eqnarray}
\label{eqyy}
  \i \dot y=(A^++P^+(\omega t))y
\\
\label{A+}
A^+:=A^-+ {\rm diag}(P^-).
\end{eqnarray}
Here ${\rm diag}(P^-)$ is the diagonal matrix formed by the
diagonal elements of $P^-$, that is   ${\rm diag}(P^-):={\rm
diag}(P^{-}_{11}(\omega t),P^{-}_{22}(\omega t),P^{-}_{33}(\omega t)...)$.  
 
The new perturbation $P^+$ is given by (the explicit
dependence of $B$ on $t$ is omitted):
\begin{equation}
\label{p+}
\begin{split}
P^+:=\left\{[A^-,B]-\i\dot B +(P^--{\rm diag}(P^-))\right\}+ \qquad
\qquad 
\\
+\left(e^{-B}A^-e^{B}-A^--[A^-,B]
\right)
+\left(e^{-B}P^-e^{B}-P^-\right)-\i\left(e^{-B}\dot Be^{B}-\dot B\right).
\end{split}
\end{equation}
According to the standard procedure we subtract the mean of the
perturbation. Namely,  we write
$A^+={\rm diag}(\lambda^+_i+\mu^+_i(\omega t))$ where $\displaystyle
\lambda^+_i=\lambda^-_i+\overline{P_{ii}(\phi)}$ (the overline denotes
angular average).  Hence the functions
$\mu^+(\phi)$ have zero average; the quantities $\lambda_i^+$ are independent
of
$\phi$ and by A3 fulfill the estimate 
$|\lambda_i^+-\lambda_i^-|\leq C_{\mu}^-i^{\delta}$.

The main step of the
proof is to construct  $B$ so as to make the curly bracket in (\ref{p+})
vanish, i.e. to solve for the unknwon $B$ the equation
\begin{equation}
\label{hom}
[A^-,B]-\i\dot B +(P^--{\rm diag}(P^-))=0\ ,
\end{equation}

The procedure explained in the previous section has to be modified
since now the eigenvalues of $A^-$ depend also on the angles
$\phi$. The construction is based on a lemma by Kuksin \cite{kuk97} that
we now summarize.

On the $n$--dimensional torus consider the equation 
\begin{equation}
\label{kuke}
-\i\sum_{k=1}^{n}\omega_k\frac\partial{\partial \phi_k} \chi(\phi)+E_1
\chi(\phi)+ E_2 h(\phi)\chi(\phi) =b(\phi)\ .
\end{equation}
Here $\chi$ denotes the unknown, while $b$,  $h$ denote given analytic
functions on $\toro^n_s$. $h$ has zero average; $E_1,E_2$ are positive
constants and $\norma h_{s}\leq1$.  Concerning the frequency vector 
$\omega=(\omega_1,...,\omega_n)$ the assumptions are:
\begin{equation}
\label{diok}
|\omega\cdot k|\geq \frac{\gamma_2}{|k|^{\tau}}\ ,\perogni
k\in\Za^n-\{0\}\ ,
\quad
|\omega\cdot k+E_1|\geq \frac{\gamma_1}{1+|k|^{\tau}}\ ,\perogni
k\in\Za^n\ .
\end{equation}
The final hypothesis is an order assumption on the magnitude of the
different parameters, namely: given $0<\theta<1$ and $C>0$ we assume
\begin{equation}
\label{E.1}
E_1^\theta\geq CE_2
\end{equation}

\begin{lemma}
\label{kuk}
(Kuksin) Under the above assumptions equation (\ref{kuke}) has a unique
analytic solution $\chi$ which for any  $0<\sigma<s$ fulfills the
estimate
\begin{equation}
\label{estk}
\norma{\chi}_{s-\sigma}\leq C_1\frac{1}{\gamma_1\sigma^{a_1}}
\exp\left(\frac{C_2}{\gamma_2^{a_2}\sigma^{a_3} }\right)\norma{b}_{s}.
\end{equation}
Here $a_1,a_2,a_3,C_1,C_2$ constants independent of
$E_1,E_2,\sigma,s,\gamma_1,\gamma_2 ,\omega$.
\end{lemma}

To apply this lemma to the construction  and  estimation of $B$, 
denote  $\G$ the Banach space of all bounded operators $B$ in $\ell^2$  such
that 
$A^{-\delta/d}BA^{\delta/d}$ extends to a bounded linear operator. The norm
in $\G$ is
 denoted
\begin{equation}
\label{G}
\norma{B}^{\G}:=\max\left\{\|B\|_0,\|A^{-\delta/d}BA^{\delta/d}\|_0
\right\} .
\end{equation}
 Moreover for the $s-$ norms of an analytic function
on the torus taking values in $\G$ (possibly 
Lipschitz-continuous  on $\omega\in\Pi^-$) we will use the notations
$$
\norma{B}^{\G}_s\ ,\qquad \norma{B}^{\G,\L}_s\ .
$$

{\it In what follows  the notation $a\sleq b$ stands for 
``there exists a constant $C$ independent of
$C_\omega^{\pm},C_{\mu}^{\pm},\gamma^{\pm}, s,\sigma,i,j,K$ (some of these
parameters will be defined later on) such that $a\leq
Cb$. Equivalently we will use the notation $b\sgeq a$.}

\begin{lemma}
\label{kuk1}  
 Let  
$\frac\delta{d-1}<\theta<1$, $\gamma_* >0$, 
$C_{\omega}^*>0$, and  $C^*>0$ be fixed.  Assume that
\begin{equation}
\label{c*}
C^*>\frac{C_\mu^-}{C_\lambda^-}\ ,\quad\gamma\geq\gamma_*\ ,
\quad C_\omega\leq C_\omega^*\ .
\end{equation}
Then for any  $0<\sigma<s$ equation (\ref{hom}) has a unique
solution $B\in\G$ analytic on $\toro^{n}_{s-\sigma}$, fulfilling the estimate
\begin{equation}
\label{sthom}
\norma{B}^{\G,\L}_{s-\sigma}
\sleq
\frac{1}{\sigma^{b_1}}
\exp\left(\frac c{\sigma^{b_2}} \right)\norma{P^-}^{\L}_{\delta,s} .
\end{equation} 
Here $c,b_1,b_2$ are constants depending only
$\theta,n,\tau,\delta, C^*,\gamma_*,C^*_\omega$.  
\end{lemma}

\proof Taking  matrix elements among eigenvectors of $A^-$, equation
(\ref{hom}) becomes
\begin{equation}
\label{comp}
-\i\sum_{k=1}^{n}\omega_k\frac{\partial}{\partial \phi_k} B_{ij}+
(\lambda_i^--\lambda_j^-)
B_{ij}+(\mu_i^-(\phi)-\mu_j^-(\phi))B_{ij}=P_{ij}, \quad i\neq j 
\end{equation}

The first inequality of (\ref{c*}) ensures that (\ref{E.1}) holds with a
suitable $C$ independent of all the relevant constants. Then a direct 
application of Kuksin's Lemma yields that (\ref{hom}) has a unique analytic
solution fulfilling the estimate
\begin{equation}
\label{sthom1}
\norma{B_{ij}}_{s-\sigma}
\sleq
\frac{1}{\gamma|i^d-j^d|}\frac1{\sigma^{a_1}}
\exp\left(\frac c{\gamma^{a_2}\sigma^{a_3}} \right)\norma{P_{ij}}_{s}
\end{equation} 
To  estimate of the sup norm of $B$ we use Lemma
\ref{unb}. To this end, first remark that $|i^d-j^d|\geq
|i-j|(i^\delta+j^\delta)$. Then consider the infinite matrices 
 of elements 
$$
\frac{P_{ij}}{(i^\delta+j^\delta)}\ ,\quad\frac{P_{ij}}{j^\delta} 
\frac{i^\delta}{(i^\delta+j^\delta)} 
$$
Assumption H2  entails a fortiori that these  infinite matrices
represent  bounded operators in
$\ell^2$. Then Lemma
\ref{unb}  yields the 
 estimate of the sup norm of $B$ and of $A^{-\delta/d}BA^{\delta/d}$,
 i.e. one has
\begin{equation}
\label{stimaG}
\norma{B}_{s-2\sigma}^\G
\sleq
\frac{1}{\sigma^{a_1+n}}
\exp\left(\frac c{\sigma^{a_3}} \right)\norma{P^-}_{\delta,s}\ 
\end{equation}
after redefinition of  $\sigma$ as $2\sigma$ and of the constant $c$. To
obtain the estimate of the Lipschitz norm we proceed as
 follows. Given a function $B$ of $\omega$ set
\begin{equation}
\label{DeltaB}
\Delta B:=B(\omega)-B(\omega'). 
\end{equation}
Applying the operator
 $\Delta$ to (\ref{comp}) one gets that $\Delta B_{ij}$
 fulfills an analogous equation. Hence 
 by  Kuksin's Lemma   its solution
 $\Delta B$ can be estimated by the same argument applied in estimating $B$. 
Dividing
 by
$|\omega-\omega'|$ and applying again Lemma
 \ref{unb} one gets
$$
\norma{\frac{\Delta
B}{\Delta\omega}}_{s-3\sigma}\sleq\left[\norma{P}^{\L}_{\delta,s}
+\frac1{\sigma^{a_1}} \exp\left(\frac c{\sigma^{a_3}}
\right)\norma{P^-}^{\L}_{\delta,s}\right]
$$
whence the proof redefining $\sigma$ as $3\sigma$ and taking the
sup as above.\quadratino

We are now ready to state and prove the main result of this section. 

\begin{lemma}
\label{+1}
Consider the system (\ref{eq}) within the stated assumptions. Assume
furthermore that also (\ref{c*}) holds. Then there exists an
anti-selfadjoint operator
$B\in\G$ analytically depending on $\phi\in\toro^n_{s-\sigma}$, and Lipschitz
continuous in $\omega\in\Pi^-$ such that
\begin{enumerate}
\item
$B$ fulfills the estimate (\ref{sthom});
\item
For any $\omega\in\Pi^-$ the
unitary operator $e^{B(\omega t)}$ transforms the system (\ref{eq}) into the
system (\ref{eqyy});
\item
 The new perturbation $P^+$ fulfills the estimate
\begin{equation}
\label{p++}
\normalip{P^+}_{\delta,s-\sigma}\sleq 
\left(\normalip{P^-}_{\delta,s}\right)^2
\exp\left(\frac
c{\sigma^{b_1}}\right)
\end{equation}
\item 
For any positive $K$ such that $\displaystyle (1+K^\tau)<
\frac{\gamma^-}{\norma{P^-}_{\delta,s}}$, there exists a closed set
$\Pi^+\subset\Pi^-$ and a  $d_4>1$ (independent of $K$)
fulfilling
\begin{equation}
\label{pi+}
\left|\Pi^--\Pi^+\right|\sleq \gamma^-\left(1+\frac1{K^{d_4}
}\right) 
\end{equation}
\item 
If $\omega\in\Pi^+$ then assumptions H1-H3 above
are fulfilled also by $A^+$ provided the constants are replaced by the new
ones  defined by
\begin{equation}
\label{new}
\gamma^+=\gamma^--\norma{P^-}_{\delta,s}(1+K^\tau)\ ,\quad C_\mu^+=
C_\mu^-+\norma{P^-}_{\delta,s}\ ,
\end{equation}
\begin{equation}
\label{new1}
C_\omega^+=C_\omega^-+\normalip{P^-}_{\delta,s}\ ,\quad
C_\lambda^+=C_\lambda^- -2\norma{P^-}_{\delta,s}\
.
\end{equation}
\end{enumerate}
\end{lemma}

\proof The estimates on $B$ are an obvious consequence of Lemma
\ref{kuk1} above. The estimate (\ref{p++}) is an immediate consequence
of Lemmas \ref{sti} and \ref{stia}. Concerning (\ref{new}) and
(\ref{new1}) the only nontrivial fact to be proved is the existence of
a set $\Pi^+$ such that, for $\omega\in\Pi^+$ (\ref{dio1}) and
(\ref{dio2}) are fulfilled with the new value of $\gamma$. Since
(\ref{dio1}) obviously holds, we examine (\ref{dio2}). First remark
that one has
$$
|\lambda_i^--\lambda_i^+|\leq\norma{P^-}_{\delta,s}i^\delta\ ;
$$
therefore, for  $|k|\leq K$ we can write, by (\ref{dio2}) and the inequality 
 $|i^d-j^d|\geq(i^\delta+j^\delta)$:
\begin{equation*}
\begin{split}
\left|\lambda_i^+-\lambda_j^+-\omega\cdot k\right|\geq
\left|\lambda_i^--\lambda_j^--\omega\cdot k\right|-
\norma{P^-}_{\delta,s}(i^\delta+ j^\delta)
\\
\geq
\frac{\gamma^--\norma{P^-}_{\delta,s}(1+K^\tau)} {1+|k|^\tau}|i^d-j^d|\ .
\end{split}
\end{equation*}
 Hence (\ref{dio2}) is
 satisfied for such values of $k$. Fix $i,j,k$ and
set:
\begin{eqnarray}
\label{rijk}
\R_{ijk}\left(\alpha\right):=\left\{\omega\in\Pi\ :\
\left|\lambda_i^+-\lambda_j^+-\omega\cdot k\right|\leq
\alpha
\right\}\ 
\\
\Pi^+:=\Pi^--\bigcup_{|k|\geq
K}\R_{ijk}\left(\frac{\gamma|i^d-j^d|}{1+|k|^\tau}\right) \ .
\end{eqnarray}
By Lemma \ref{mis1} the set (\ref{rijk}) is nonempty only if $|k|\geq
|i^d-j^d|(C^-_\lambda-\gamma^-)$,  and 
by Lemma 
\ref{meas}, one has
$$
\left|\R_{ijk}\left(\frac{\gamma|i^d-j^d|}{1+|k|^\tau}\right)\right|
\sleq\frac{\gamma|i^d-j^d|}{(1+|k|^\tau)|k|}\ . 
$$
Since $|i^d-j^d|\geq|i-j|(i^{d-1}+j^{d-1})$, the cardinality of the
set $\{ (i,j)\;|\; |i^d-j^d|\leq L\}$ is bounded by an absolute
constant times $L^{2/(d-1)}$. Hence if $\tau>n+2/(d-1)$ one has
\begin{equation}
\begin{split}
\left|\bigcup_{ijk:|k|\geq
K}\R_{ijk}\left(\frac{\gamma|i^d-j^d|}{1+|k|^\tau}\right) \right|\sleq
\sum_{|k|\geq K,|i^d-j^d|\leq C|k|}
\frac{\gamma|i^d-j^d|}{(1+|k|^\tau)|k|}
\\
 \sleq \gamma\sum_{s\geq
K}\frac1{s^{\tau-n +1-2/(d-1)}}\sleq \frac\gamma {K^{d_4}}\ ,
\end{split}
\end{equation}
and this proves the assertion.\quadratino
\section{Iteration}
\setcounter{equation}{0}%
\setcounter{theorem}{0}%
\setcounter{proposition}{0}%
\setcounter{lemma}{0}%
\setcounter{corollary}{0}%
\setcounter{definition}{0}%
\setcounter{remark}{0}%
\setcounter{example}{0}%

In this section we set up the iteration needed to prove the stated results.
First we preassign the values of the various constants occurring in the
iterative estimates. Hence we keep  $\epsilon$, $K$, $s$ and $\gamma$
fixed and define, for $l\geq1$, 
\begin{equation}
\epsilon_l:=\epsilon^{(4/3)^{l}}\ ,\quad
\sigma_l:=\frac{s}{4l^2} \ ,\quad s_l=s_{l-1}-\sigma_l  \ ,\quad
K_l:=l K 
\end{equation}
\begin{equation}
\gamma_l=\gamma_{l-1}-4\epsilon_{l}(1+K_l^\tau)  \ ,\quad
C_{\mu,l}=C_{\mu,l-1} +\epsilon_l\ ,\quad
\end{equation}
\begin{equation}
C_{\lambda,l}=C_{\lambda,l-1} -2\epsilon_l\  ,\quad
C_{\omega,l}=C_{\omega,l-1} +\epsilon_l\ .
\end{equation}
The initial values of the sequences are chosen as follows:
$$
\gamma_0:=\gamma\ ,s_0=s, \quad C_{\mu,0}:=0\ ,\quad
C_{\lambda,0}:=C_{\lambda}\ ,\quad  C_{\omega,0}:=0\ .
$$

\begin{proposition}
\label{l+1}
There exist  $\epsilon_*=\epsilon_*(\gamma)>0$ and, for any $l\geq1$,
a closed set $\Pi_l^\gamma\subset\Pi$ such that, if
$|\epsilon|<\epsilon_*$, one can construct for
$\omega\in\Pi^{\gamma}_l$  
 a unitary transformation $U^l_{\epsilon}$,
 analytic and quasiperiodic in $t$  with
frequencies $\omega$,  mapping the system (\ref{eq0}) into the
system
\begin{equation}
\label{l+sys}
\i \dot x=(A^l+P^l(\omega t))x
\end{equation}
where:
\begin{enumerate}
\item $U^l_{\epsilon}(\omega t)$ is as follows: $U^l_{\epsilon}(\phi)
=\e^{B^1_{\epsilon}(\phi)}\e^{B^2_{\epsilon}(\phi)}...\e^{B^l_{\epsilon}
(\phi)}$, and the anti-selfadjoint operators $B^j_{\epsilon}\in\G$,
j=1,...,l depend analytically on $\phi\in\toro^n_{s-\sigma_l}$, are
Lipschitz continuous in $\omega\in\Pi^\gamma_l$ and fulfill
(\ref{sthom}) with $P_{l-1}$, $\sigma_l$ in place of $P^-, \sigma$,
respectively.
\item
 $A^l$ has the form of (\ref{A}) with the upper index  ``minus''
replaced by $l$, i.e.
\begin{equation}
\label{Al}
A^l={\rm diag}(\lambda^l_1(\om)+\mu_1^l(\omega
t,\om),\lambda^l_2(\om)+\mu_2^l(\omega t,\om),\lambda^l_3(\om)+\mu_3^l(\omega
t,\om),...)\ ,
\end{equation} 
\item The corresponding $\lambda_i^l$ and
$\mu_i^l$ fulfill conditions H1, H2, H3 of the previous section, provided
$\lambda^-_i, \mu^-_i$ are replaced by $\lambda^l_i, \mu^l_i$, respectively.
\item  
The following estimates hold
\begin{equation}
\label{stimeP}
\norma{P^l}_{\delta,{s_l}}\leq\epsilon_l\ ,\quad
\norma{B^l_\epsilon}^{\G,\L}_{\delta,s_{l+1}} \leq\epsilon_l\ ,\quad
\left|\Pi_l^\gamma-\Pi_{l+1}^\gamma   \right|\leq
\gamma_l\left(1+\frac1{(lK)^{d_4}} \right) .
\end{equation}
\end{enumerate} 
\end{proposition}

\proof We proceed by induction applying Lemma
\ref{+1}. First we want to apply it to the original system
(\ref{eq0}) to the effect of obtaining a system of the form (\ref{l+sys})
with
$l=1$. To this end remark that (\ref{eq0}) satisfies all the assumptions of
Lemma
(\ref{+1}) except the nonresonance conditions (\ref{dio1}) and (\ref{dio2})
on the frequencies. We have to restrict the set of the
frequencies. Define therefore
$$
\Pi_0^\gamma:=\Pi-\bigcup_{ijk}\R_{ijk}\left(\frac{\gamma|i^d-j^d|}
{1+|k|^\tau} \right)
$$
and  remark that, by Lemma \ref{meas},  
$\left|\Pi-\Pi_0^{\gamma}\right|\sleq \gamma\ .$
Hence we can apply Lemma \ref{+1} and  the starting point of our
induction procedure is established.

To go from step $l$ to step $l+1$ one has to verify that the
assumptions of Lemma \ref{+1} are satisfied for any $l$. More
specifically, defining $\gamma^*:=\gamma/2$ and fixing $C^*$ and
$C^*_\omega$ we must verify that (\ref{c*}) holds. It is easy to check 
that this is true provided $\epsilon$ is smaller than a constant which in
particular vanishes as $\gamma\to 0$. Then it is immediately
 realized that the conclusions of Lemma \ref{+1} imply the thesis if
$\epsilon$ is small enough (independently of $l$).  \quadratino

\vskip
5pt\noindent {\bf Proof of Theorem \ref{main} } Proposition \ref{l+1}
ensures the existence of $\epsilon^*>0$ such that, for
$|\epsilon|<\epsilon^*(\gamma)$,
$\lim_{l\to\infty}\gamma_l=\gamma^\infty,$ 
$\gamma^\infty>\gamma/2$, and $\lim_{l\to\infty}s_l=s/2$.  This entails
the uniform convergence of the operator valued sequence of functions
$U^l_\epsilon$ on $\toro^n_{s/4}$.  Hence the limit, denoted
$U^{\infty}_{\epsilon}(\om t)$, will be analytic and
quasi-periodic. Moreover, writing $A_\infty:= {\rm
diag}(\lim_{l\to\infty}(\lambda_i^l+\mu_i^l))$, one has
$$
\lim_{l\to\infty}\norma{A_l(\phi)-A_{\infty}(\phi)}_{\delta}=0\ 
$$
uniformly on $\toro^n_{s/4}$.
This proves T1 and T2. The first two estimates of T3 are also clearly
implied by the above convergence. Set now $\displaystyle
\Pi^{\gamma}=\bigcap_{l=1}^{\infty}\Pi^{\gamma/2}_l.\,$ By the second of
(\ref{stimeP}) we have
$$
|\Pi-\Pi^{\gamma}|\sleq \gamma_0=\gamma
$$
Denote now  $\gamma(\epsilon^*)$ the inverse function of 
$\gamma\mapsto\epsilon^*(\gamma)$, and define
$\Pi^\epsilon:=\Pi^{\gamma(\epsilon)}$. Then the third estimate of
assertion T3 follows. \quadratino

\vskip 5pt\noindent {\bf Proof of
Corollaries \ref{cor1} and \ref{mainc} } Integration of
(\ref{diagonale}) yields:
$$
\chi_i(t)= \chi_i(0)e^{\i\lambda_i^{\infty}t}e^{iF_i^{\infty}(t)}, \qquad 
F_i^{\infty}(t):=\sum_{k\in\Za^n-\{0\}}\frac{\mu_{i,k}^{\infty}}{
\om\cdot k t }(e^{\i\om\cdot k}-1), \quad i=0,1,\ldots
$$
where $\mu_{i,k}^{\infty}, k=\Za^n$, are the Fourier coefficients of
$\mu_i(\phi)$. Setting $x_i:=\e^{\i F_i^{\infty}(t)}\chi_i$ we get
$\i\dot{x}_i=\lambda_i^{\infty}x_i$. Formula  (\ref{psi}) follows taking $\chi
=U_{\epsilon}\phi$. Moreover it is trivially verified that   
$\displaystyle \phi_i^0(\omega t)e^{i\lambda^{\infty}_i t}$ solves
(\ref{I0}) if and only if 
$\lambda^{\infty}_i+\langle k,\omega\rangle$ is an eigenvalue of
(\ref{flo}).
\quadratino\vskip 5pt\noindent
 {\bf Proof of Theorem \ref{Floquet} } Let $A$ denote the maximal
operator in $L^2(\Re)$ generated by the differential expression $\displaystyle
-\frac{d^2}{dx^2}+Q(x)$. It is well known that $A$ is self-adjoint,
strictly positive and has compact resolvent and that, denoting $\lambda_i,
i=1,2,\ldots$ its eigenvalues, one has 
$\displaystyle 
\lambda_i\sim i^{\frac{2\alpha}{\alpha +2}}, i\to\infty.$
Hence   condition A1 is fulfilled if $\alpha>2$. 
$A$ can be realized also as a pseudifferential operator of symbol
$\displaystyle
\sigma_{A}(x,\xi):=\xi^2+Q(x)$ under Weyl quantization. 
$\sigma_{A}(x,\xi)$ belongs to the symbol class
$\Gamma^{\alpha}_{\rho}(\Re):=\Gamma^{\alpha}_{\rho}$ for any 
$0<\rho<1$ (notations as in \cite{Sh}, Sect.23). This class of symbols generates
the class $G^{\alpha}_{\rho}$ of pseudodifferential operators in $L^2(\Re)$
under the Weyl quantization formula:
$$
(Au)(x)=\frac{1}{2\pi^n}\int_{\Re^n\times\Re^n}e^{i(x-y)\xi}
\sigma_A(\frac{x+y}
{2},\xi)u(y)\,dyd\xi, \qquad u\in\S(\Re)
$$
 The inverse
$[A+1]^{-1}$, whose principal symbol is
$\displaystyle \sigma_{(A+1)^{-1}}(x,\xi)=(\xi^2+Q(x)+1)^{-1}$, belongs to
the the  class $G^{-\alpha}_{\rho}$.  The functional calculus for
pseudodifferential operators (see e.g.\cite{Sh}, Chapt.II.10,11 or 
\cite{DiSj}, Chapt.8) can be applied to operators in these classes. Hence the
self-adjoint operator
$A^{q}, q>0$ defined by the spectral theorem can also be realized a
pseudodifferential operator in $G^{\alpha q}_{\rho}$, with symbol in
$\Gamma^{\alpha q}_{\rho}$. Its principal symbol is
$\displaystyle
\sigma_{A^q}(x,\xi):=(\xi^2+Q(x))^q$, and the principal symbol of
$\displaystyle [A^q+1]^{-1}\in G^{-\alpha q}_{\rho}$ is $\displaystyle
\sigma_{(A^q+1)^{-q}}(x,\xi):=[(\xi^2+Q(x))^q+1]^{-1}$. By assumption  the
symbol of the perturbation $V$ belongs to $\Gamma^{\beta}_{\rho}$ for any 
$0<\rho<1$, and hence $V$ belongs to $G^\beta_{\rho}$. By the composition
property, the operator 
$T:=V[A^q+1]^{-1}$ admits a symbol in $\Gamma^{-\alpha q+\beta}_\rho$,
and it will be bounded if $-\alpha q+\beta \leq 0$ (\cite{Sh}, Thm.
24.3). In turn, it is enough to verify this property for the principal
symbol, which in this case, by the composition formula, is given by
$$
\sigma^P_T(x,\xi)=v(x,\xi;\phi) [(\xi^2+Q(x))^q+1]^{-1}.
$$
Since here $q=\delta/d$, $|\sigma^P_T(x,\xi)|$ is bounded
$\forall\,(x,\xi)\in\Re^n\times\Re^n$ if there is $D>0$ such that
$\displaystyle |v(x,\xi;\phi)|\leq D(\xi^2+|x|^{\alpha})^{\delta/d}$. If
$V\sim |x|^{\beta}$ as $|x|\to\infty$ the inequality is satisfied for $\beta
\leq\alpha\delta/d$. Now we can set $\displaystyle
1<d=\frac{2\alpha}{\alpha+2}$. Then $\delta<d-1$ means $\displaystyle 0<\delta<
\frac{\alpha -2}{\alpha+2}$ and therefore $\displaystyle \beta < \frac{\alpha
-2}{2}$.
\quadratino

\section{Technical Lemmas}
\setcounter{equation}{0}%
\setcounter{theorem}{0}%
\setcounter{proposition}{0}%
\setcounter{lemma}{0}%
\setcounter{corollary}{0}%
\setcounter{definition}{0}%
\setcounter{remark}{0}%
\setcounter{example}{0}%
\miolemma{B.3}{Let $f_j$ be analytic functions on $\toro^n_s$. Then 
for any  $0<\sigma< s$ one has
$$
\Big(\sum_{j\geq1}\norma{f_j}_{s-\sigma}^2\Big)^{1/2}\leq\frac{4^n}
{\sigma^n}\|{\Big(\sum_{j\geq1}\left|f_j\right|^2\Big)^{1/2}}\|_s
$$
 }

\proof This is Lemma B.3 of \cite{KP};  we
reproduce  its proof here for convenience of the reader. First consider the
case $n=1$. For each $j\geq 1$ there exists a point
$\phi_j\in\toro_{s-\sigma}$ such that 
$$
\norma{f_j}_{s-\sigma}\leq|f_{j}(\phi_j)|\ .
$$ 
By the Cauchy integral formula
$$
f_j(\phi_j)=\frac1{2\pi\i}\int_{\partial\Gamma_\rho}\frac{f_j(\zeta)}{\zeta-\phi_j}
d\zeta\ ,
$$
where $0<\rho<\sigma$, is a parameter independent of $j$, and
$\partial\Gamma_\rho$ is the boundary of  the set $\Gamma_\rho:=\left\{\phi \
:\ -\rho<{\rm Re}\phi<2\pi+\rho\ ,\
-(s-\sigma+\rho)<{\rm Im}\phi<s-\sigma+\rho\right\}$. One has
\begin{equation}
\begin{split}
\Big(\sum_{j\geq1}\norma{f_j}_{s-\sigma}^2\Big)\leq\Big(\sum_{j\geq1}
\Big|\frac1{2\pi\i}\int_{\partial\Gamma_\rho}\frac{f_j(\zeta)}{\zeta-\phi_j}
d\zeta
\Big|^2\Big)^{1/2}
\\
\leq \frac{1}{2\pi}\int_{\Gamma_\rho}\Big(\sum_{j\geq1}
\Big|\frac{f_j(\zeta) 
}{\zeta-\phi_j}\Big|^2\Big)^{1/2}|d\zeta|
\leq\frac4\rho\sup_{\toro_s}\Big(\sum_{j\geq1}|f_j(\phi)|^2
\Big)^{1/2}\ 
.
\end{split}
\end{equation}
Taking the limit $\rho\to\sigma$ one gets the result. The case $n>1$
follows similarly.\quadratino

\begin{lemma}
\label{unb}
Let $F=(F_{ij})$ be a bounded operator on
$\ell^2$, and let the matrix elements $(F_{ij})$ be analytic functions of
$\phi\in\toro^n_s$. Let $R=(R_{ij})$ be another operator with matrix
elements depending analytically on $\phi\in\toro^n_\sigma$ and such that
$$
\sup_{\phi\in\toro^n_s}|R_{ij}(\phi)|\leq \frac1{|i-j|}{\sup_{\phi\in\toro^n_s}
|F_{ij}(\phi)|}
\ ,\quad i\not=j\ .
$$
Then, for any $\phi\in\toro_s^n$, $R$ is bounded in $\ell^2$ and for any
positive
$\sigma<s$ 
it fulfills the estimate
$$
\norma{R}_{0,s-\sigma}\leq \frac{4^{n+1}}{\sigma^{n}}\norma{F}_{0,s}\ .
$$
\end{lemma}
\proof This is Lemma
 B.4 of \cite{KP}; again we reproduce its proof here for convenience of the
reader.  Fix
$\phi\in
\toro_{s-\sigma}$. By Lemma \ref{B.3} and the Schwarz inequality we have
\begin{equation}
\begin{split}
\sum_{j\geq1} \left|R_{ij}(\phi)\right|\leq
\sum_{j\geq1}\norma{R_{ij}}_{s-\sigma } \leq
\Big(\sum_{j\geq1}\norma{F_{ij}}_{s-\sigma}^2\Big)^{1/2}\Big(
\sum_{j\not=i}\frac1{|i-j|^2} \Big)^{1/2} 
\\ 
\leq
\frac{4^{n+1}}{\sigma^n}\sup_{\toro_{s}^n} \Big(\sum_{j\geq1}|
F_{ij}|^2\Big) ^{1/2} 
\leq \frac{4^{n+1}}{\sigma^n}\norma{F}{0,s}\ .
\end{split}
\end{equation}
The same estimate holds for $\sum_{i\geq1}|F_{ij}(\phi)|$. Hence, for
$\phi\in\toro^{n}_\sigma$ 
\begin{equation}
\begin{split}
\norma{R(\phi)v}^2=\sum_{i\geq1}\Big(\sum_{j\geq
1}|R_{ij}(\phi)||v_j| \Big)^2
\leq 
\sum_{i\geq1}\Big(\sum_{j\geq
1}|R_{ij}(\phi)| \Big)\Big(\sum_{j\geq
1}|R_{ij}(\phi)||v_j|^2 \Big)
\\
\leq \Big(\sum_{j\geq
1}|R_{ij}(\phi)| \Big)\Big(\sum_{i\geq
1}|R_{ij}(\phi)| \Big)\Big(\sum_{j\geq1}|v_j|^2\Big)
\leq \Big(\frac{4^{n+1}}{\sigma^n}\norma{F}_{0,s}\Big)^2\norma v^2
\end{split}
\end{equation}
which proves the result.\quadratino

\miolemma{sti}{Let $B\in\G$ be a bounded  anti-selfadjoint
operator, and let $P\in\B^{\delta}$ be a selfadjoint operator. Then one
 $e^{-B}Pe^B\in\B^\delta$ and, provided $\norma{B}^{\G}\leq 1/2$,
the following estimate holds
\begin{equation}
\label{stB}
\norma{e^{-B}Pe^B-P}_{\delta}\leq 4\norma{P}_{\delta}\norma{B}^{\G}
\end{equation}
Moreover, if both $B$ and $P$ are  Lipschitz continuous with respect to 
$\omega\in\Pi$,  then
\begin{equation}
\label{stlip}
\normalip{e^{-B}Pe^B-P}_{\delta}\leq
4\normalip{P}_{\delta} \norma{B}^{\G,\L}
\end{equation}
}
\proof Define $P(t):=e^{-tB}Pe^{tB}$. Then $P(t)$  fulfills the
linear differential equation
$$
\dot P=[B,P]\ ,\quad P(0)=P
$$
whence
$$
\|\dot P(t)\|_{\delta} \leq 2\norma{B}^{\G}\norma{P(t)}_{\delta}\
\Longrightarrow 
\norma{P(t)}_{\delta} \leq\exp\left(2\norma{B}^{\G}t\right)\norma{P}_
{\delta}\ . 
$$
Then  (\ref{stB}) follows on account of
$$
P(t)-P=\int_0^t[B,P(s)]ds \ .
$$
To obtain the Lipschitz estimate
remark that (same notation as in the proof of Lemma
\ref{kuk1}), $\Delta P$ fulfills the
equation
$$
(\Delta P)\dot{\ }=[\Delta B,P]+[B,\Delta P]\ ,
$$
and then  proceed as in the estimation of the operator norm.
\quadratino

\miolemma{stia}{Let $B\in\G$ be the solution of  equation
(\ref{hom}) and let $0<\sigma<s/2$. Then: 
\begin{eqnarray}
\norma{e^{-B}A^-e^{B}-A^--[A^-,B]}_{\delta,{s-2\sigma}}\sleq
\norma{B}^{\G}_{s-\sigma}\left(\frac1\sigma\norma{B}_{\delta,{s-\sigma}}+
\norma
{P^-}_{\delta}\right) 
\\
\normalip{e^{-B}A^-e^{B}-A^--[A^-,B]}_{\delta,{s-2\sigma}}\sleq
\norma{B}^{\G,\L}_{s-\sigma}\left(\frac1\sigma\normalip{B}_{\delta,{s-\sigma}}+
\normalip
{P^-}_{\delta}\right) 
\end{eqnarray}
}
\proof The proof goes by the same argument of Lemma \ref{sti}; just
use the formula
$$
e^{-B}A^-e^{B}-A^--[A^-,B]=\int_0^1ds\int_0^se^{-s_1B}[[A^-,B],B]  e^{s_1B}ds_1
$$
and compute $[A^-,B]$ from equation (\ref{hom}). The the assertion  easily
follows. \quadratino

\miolemma{mis1}{Assume that the sequence $\lambda_i$ fulfills
Assumption H1 of Sect.2 and equation \ref{lip} and fix
$\alpha<C_\lambda/2$; then the set $\R_{ijk}(\alpha|i^d-j^d|)$ is
empty if $|k|<(C_\lambda/2)|i^d-j^d|$.} 

The proof of this Lemma is straigthforward and  therefore omitted. 

\miolemma{meas}{If the sequence $\lambda_i$ fulfills
assumption H1)  and  (\ref{lip}) 
 $\exists\,C>0$ such that, if 
$$
\frac{nC_\omega}{C_\lambda}\leq\frac12
$$
then one has 
$$
|\R_{ijk}(\alpha)|\leq \frac{C\alpha }{|k|}\ . 
$$
}
\proof Following the  proof of Lemma 5
of ref. \cite{poe96a} we fix $v\in\left\{-1,1\right\}^n$ such that
$v\cdot k=|k|$ and write $\omega=av+w$ with $w\in v^{\perp}$. One has
that, as afunction of $a$
\begin{eqnarray*}
(\omega\cdot k)\big\vert_s^t=|k|(t-s), 
\qquad
\left(\lambda_i-\lambda_j\right)\big\vert_s^t\leq
C_\omega(i^\delta+j^\delta) |v|(t-s)\ .
\end{eqnarray*}
so, by Lemma \ref{mis1}, either $\R_{ijk}$ is empty or
$$
(\omega\cdot k+\lambda_i-\lambda_j)\big\vert_s^t\geq|k|(t-s)\left(
1-\frac{nC_\omega2}{C_\lambda}   \right)\geq\frac12 |k|(t-s)\ ,
$$
and therefore by the assumption we can conclude
$$
\left|\R_{ijk}(\alpha)\right|\leq\frac4{|k|}\alpha\ .
$$
\quadratino

\vfill\eject

\end{document}